\documentclass{article}
\usepackage{smc2017}
\usepackage{times}
\usepackage{ifpdf}
\usepackage[english]{babel}
\usepackage{cite}
\usepackage{makecell}
\usepackage{multirow}
\usepackage{hhline}

\def\papertitle{AUDIO-TO-SCORE ALIGNMENT OF PIANO MUSIC USING RNN-BASED AUTOMATIC MUSIC TRANSCRIPTION}
\def\firstauthor{Taegyun Kwon}
\def\secondauthor{Dasaem Jeong}
\def\thirdauthor{Juhan Nam}

\newif\ifpdf
\ifx\pdfoutput\relax
\else
   \ifcase\pdfoutput
      \pdffalse
   \else
      \pdftrue
\fi

\ifpdf 
  \usepackage[pdftex,
    pdftitle={\papertitle},
    pdfauthor={\firstauthor, \secondauthor, \thirdauthor},
    bookmarksnumbered, 
    pdfstartview=XYZ 
   ]{hyperref}

  \usepackage[pdftex]{graphicx}
  \graphicspath{{./figures/}}
  \DeclareGraphicsExtensions{.pdf,.jpeg,.png}

  \usepackage[figure,table]{hypcap}

\else 
  \usepackage[dvips,
    bookmarksnumbered, 
    pdfstartview=XYZ 
  ]{hyperref}  

  \usepackage[dvips]{epsfig,graphicx}
  \graphicspath{{./figures/}}
  \DeclareGraphicsExtensions{.eps}

  \usepackage[figure,table]{hypcap}
\fi

\hypersetup{
    colorlinks,%
    citecolor=black,%
    filecolor=black,%
    linkcolor=black,%
    urlcolor=black
}

\title{\papertitle}

\oneauthor
  {\firstauthor, \secondauthor, and \thirdauthor} {Graduate School of Culture Technology, KAIST \\ %
   {\tt \href{mailto:juhannam@kaist.ac.kr}{\{ilcobo2, jdasam, juhannam\}@kaist.ac.kr}}}

\begin{document}
\capstartfalse
\maketitle
\capstarttrue
\begin{abstract}
We propose a framework for audio-to-score alignment on piano performance that employs automatic music transcription (AMT) using neural networks.
Even though the AMT result may contain some errors, the note prediction output can be regarded as a learned feature representation that is directly comparable to MIDI note or chroma representation.
To this end, we employ two recurrent neural networks that work as the AMT-based feature extractors to the alignment algorithm. One predicts the presence of 88 notes or 12 chroma in frame-level and the other detects note onsets in 12 chroma. We combine the two types of learned features for the audio-to-score alignment.
For comparability, we apply dynamic time warping as an alignment algorithm without any additional post-processing.
We evaluate the proposed framework on the MAPS dataset and compare it to previous work. 
The result shows that the alignment framework with the learned features significantly improves the accuracy, achieving less than 10 ms in mean onset error. 
\end{abstract}

\section{Introduction}\label{sec:introduction}



Audio-to-score alignment (also known as score following) is the process of temporally fitting music performance audio to its score.  
The task has been explored for quite a while and utilized mainly for interactive music applications, for example, automatic page turning, computer-aided accompaniment or interactive interface for active music listening \cite{Arzt2008AutomaticListening,Dannenberg2006MusicAccompaniment}. 
Another use case of audio-to-score alignment is performance analysis which examines performer's interpretation of music pieces in terms of tempo, dynamics, rhythm and other musical expressions \cite{Widmer2003InFactor}. To this end, the alignment result must be sufficiently precise having high temporal resolution. 
It was reported that the just-noticeable difference (JND) time displacement of a tone presented in a metrical sequence  is about 10 ms for short notes \cite{friberg1993perception}, which is beyond the current accuracy of the automatic alignment algorithm. 
This challenge has provided the motivation for our research. 

There are two main components in audio-to-score alignment: features used in comparing audio to score, and alignment algorithm between two feature sequences. 
In this paper, we limit our scope to the feature part.
A typical approach is converting MIDI score to synthesized audio and comparing it to performance audio using various audio features. 
The most common choices are time-frequency representations through short time Fourier transformation (STFT) \cite{Dixon2005MATCH:Chest} or auditory filter bank responses \cite{Ewert2009HighFeatures}.
Others suggested chroma audio features, which are designed to minimize differences in acoustic quality between two piano audio such as timbre, dynamics and sustain effects \cite{Ewert2009HighFeatures}. 
However, the design process by hands relies on trial-and-error and so is time-consuming and sub-optimal. 
Another approach to audio-to-score alignment is converting the performance audio to MIDI using automatic music transcription (AMT) systems and comparing the performance to score in the MIDI domain \cite{Arzt2014TheCompanion}. 
The advantage of this approach is that the transcribed MIDI is robust to timbre and dynamics variations by the nature of the AMT system if it predicts only the presence of notes.
In addition, the synthesis step is not required. However, the AMT system must have high performance to predict notes accurately, which is actually a challenging task. 


In this paper, we follow the AMT-based approach for audio-to-score alignment. 
To this end, we build two AMT systems by adapting a state-of-art method using recurrent neural networks \cite{Bock2012PolyphonicNetworks} with a few modifications. 
One system takes spectrograms as input and is trained in a supervised manner to predict a binary representation of MIDI in either 88 notes or chroma.  
The prediction does not consider intensities of notes, i.e. MIDI velocity. 
Using this system only however does not provide precise alignment because onset frames and sustain frames are equally important.  
In order to make up for the limitation, we use another AMT system that is trained to predict the onsets of MIDI notes in chroma domain. 
This was inspired from Decaying Locally-adaptive Normalized Chroma Onset (DLNCO) feature by Ewert et al. \cite{Ewert2009HighFeatures}. 
Following the idea, we employ decaying chroma note onset features which turned out to offer not only temporally precise points but also make onset frames salient.
Finally, we combine the two MIDI domain features and conduct dynamic time warping algorithm on the feature similarity matrix. 
The evaluation on the MAPS dataset shows that our proposed framework significantly improves the alignment accuracy compared to previous work.

\section{SYSTEM DESCRIPTION}\label{sec:SYSTEM}

\begin{figure} 
\includegraphics[width=\columnwidth]{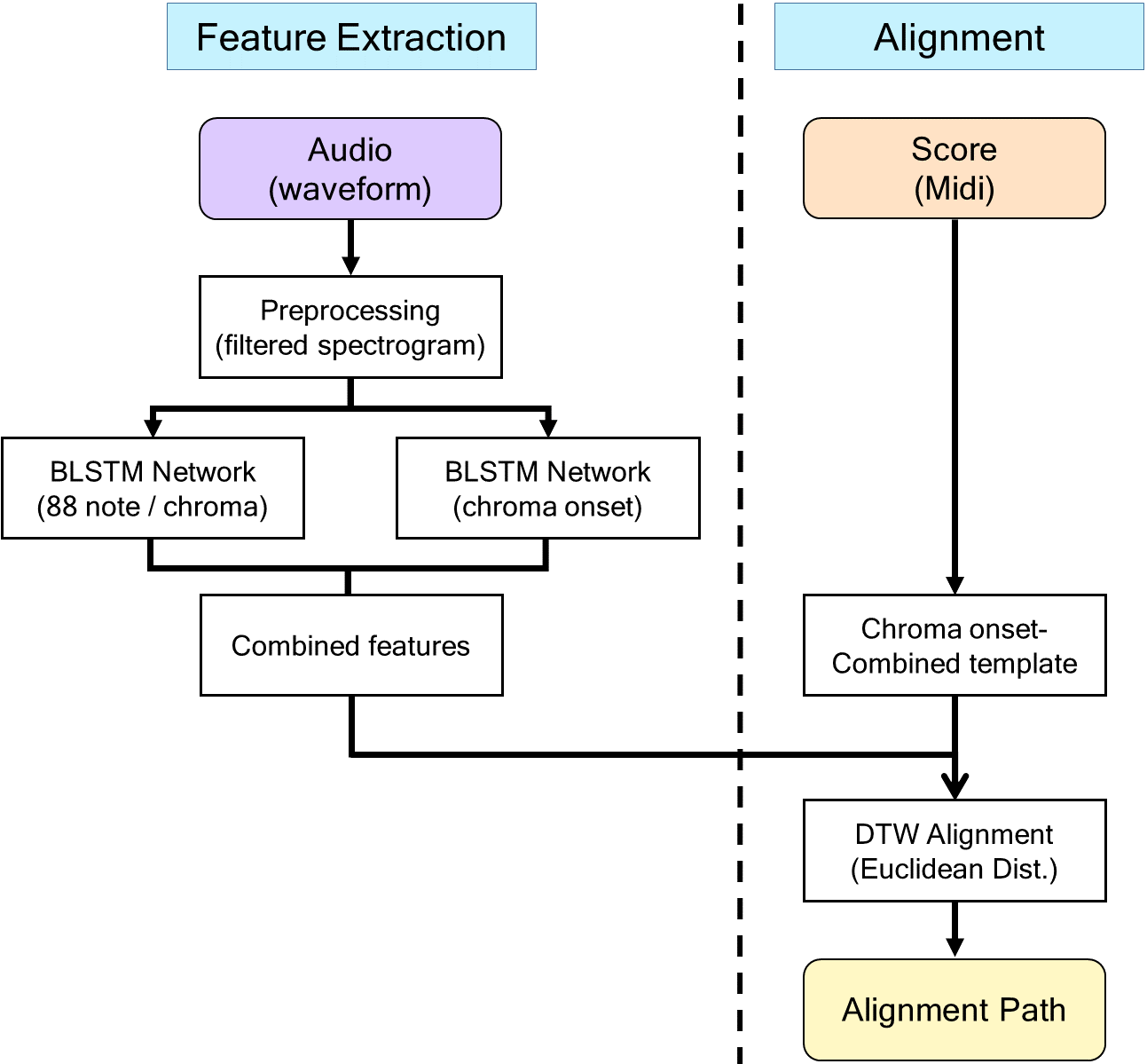}
\caption{Flow diagram of proposed audio-to-score alignment system}
\label{fig:work_flow}
\end{figure}

The proposed framework is illustrated in Figure \ref{fig:work_flow}.
The left-hand presents the two independent AMT systems that return either 88 note or chroma output and chroma onset output, respectively. The outputs are concatenated and aligned with the score MIDI through dynamic time warping (DTW).
Since our main idea is not improving the performance of AMT system but rather utilizing a neural-network based system that produces features for audio-to-score alignment, we borrowed the state-of-art AMT system proposed by B{\"o}ck and Schedl \cite{Bock2012PolyphonicNetworks}. However, we slighly modified the training setting for our purpose.


\subsection{Pre-processing}
As aforementioned, our AMT system is based on the existing model. Therefore,  we used the same multi-resolution STFT with semitone spaced logarithmic compression in the model.
It first receives audio waveforms as input and computes two types of short time Fourier transform (STFT), one with a short window (2048 samples, 46.4 ms) and the other with a long window (8192 samples, 185.8 ms), with the same overlap (441 samples, 10 ms). 
The STFT with a short time window gives temporally sensitive output while the one with a longer window offers better frequency resolution.
A Hamming window was applied on the signal before the STFT. We only take magnitude of the STFT, thereby obtaining spectrogram with 100 frames/sec.

To apply logarithmic characteristics of sound intensity, a log-like compression with a multiplication factor 1000 is applied on the magnitude of spectrograms.
We then reduce the dimensionality of inputs by filtering with semi-tone filterbanks. The center frequencies are distributed according to the frequencies of the 88 MIDI notes and the widths are formed with overlapping triangular shape.
This process is not only effective for reducing size of inputs but also for suppressing variance in piano tuning by merging neighboring frequency bins.
In the low frequency, some note bins become completely zero or linear summation of neighboring notes due to the low frequency resolution of the spectrogram.
We remove those dummy note bins, thereby having 183 dimensions in total. We augmented the input by concatenating it with the first-order difference of the semitone filtered spectrogram. We observed a significant increase of the transcription performance with this addition.

\subsection{Neural Network}
The B{\"o}ck and Schedl model uses a recurrent neural network (RNN) using the Long Short Term Memory (LSTM) architecture. Compared to feedforward neural networks, RNNs are capable of learning temporal dependency of sequential data, which is the property found in music audio. Also, the LSTM unit has a memory block updated only when an input or forget gate is open, and the gradients can propagate through memory cells without being multiplied each time step. This property enables LSTM to learn long-term dependency. In our task, the LSTM is expected to learn the continuity of onset, sustain and offset within a note as well as the relation among notes. The LSTM units are also set to be bidirectional, indicating that the input sequence is not only presented in order but also in the opposite direction. Throughout backward and forward layers together, the networks can access to both history and future of the given time frame.

While the B{\"o}ck and Schedl model used a single network that predicts 88 notes, we use two types of networks; one predicts 88 notes or 12 chroma and the other predicts 12 chroma onsets. In the 88-note network, we reduced the size to two layers of 200 LSTM units as it performed better in our experiments. In the 12-chroma, we downsized it further having 100 LSTM units on the first layer and 50 LSTM units on the second layer. On top of the LSTM networks, a fully connected layer with sigmoid activation units are added as the output layer. Each output unit corresponds to one MIDI note or chroma (i.e. pitch class of the MIDI note). 


\subsubsection{Backpropagation}

Theoretically, LSTM can learn any length of long-term dependency through backpropagation through time (BPTT) with a desired number of time steps.
In practice, it requires large memory and heavy computation because all past history of network within the backpropagation length should be stored and updated.
To overcome this difficulty, a truncated backpropagation method  \cite{Williams1990AnTrajectories} is  usually applied for long sequences (also with long time dependency).
In the truncated backpropagation, input sequences are divided into shorter sequences and the last state of each segment is transfered to the consecutive segment. 
Therefore, even though the backpropagation is only computed in each segment, it can serve as an approximation to full-length backpropagation.
For a bidirectional system, however, the backward flow requires computation on the full future and thus the truncated backpropagation requires large memory as well.
To imitate the advantage of the truncated backpropagation within the computational availability, we split the input sequence into relatively large sequences and perform full-length backpropagation within each segment. 
We conducted grid search on the segmentation length between 10 frames to 300 frames (100 to 3000 ms) and finally settled down to 50 frames (500 ms). This was long enough to catch up the continuity of individual notes and also was not computationally expensive.   
We conducted a comparative experiment between a unidirectional model with the truncated backpropagation and a bidirectional model with a non-transferred segmentation. 
The result showed that the bidirectional model performs better.

\begin{figure}[t]
\includegraphics[width=\columnwidth]{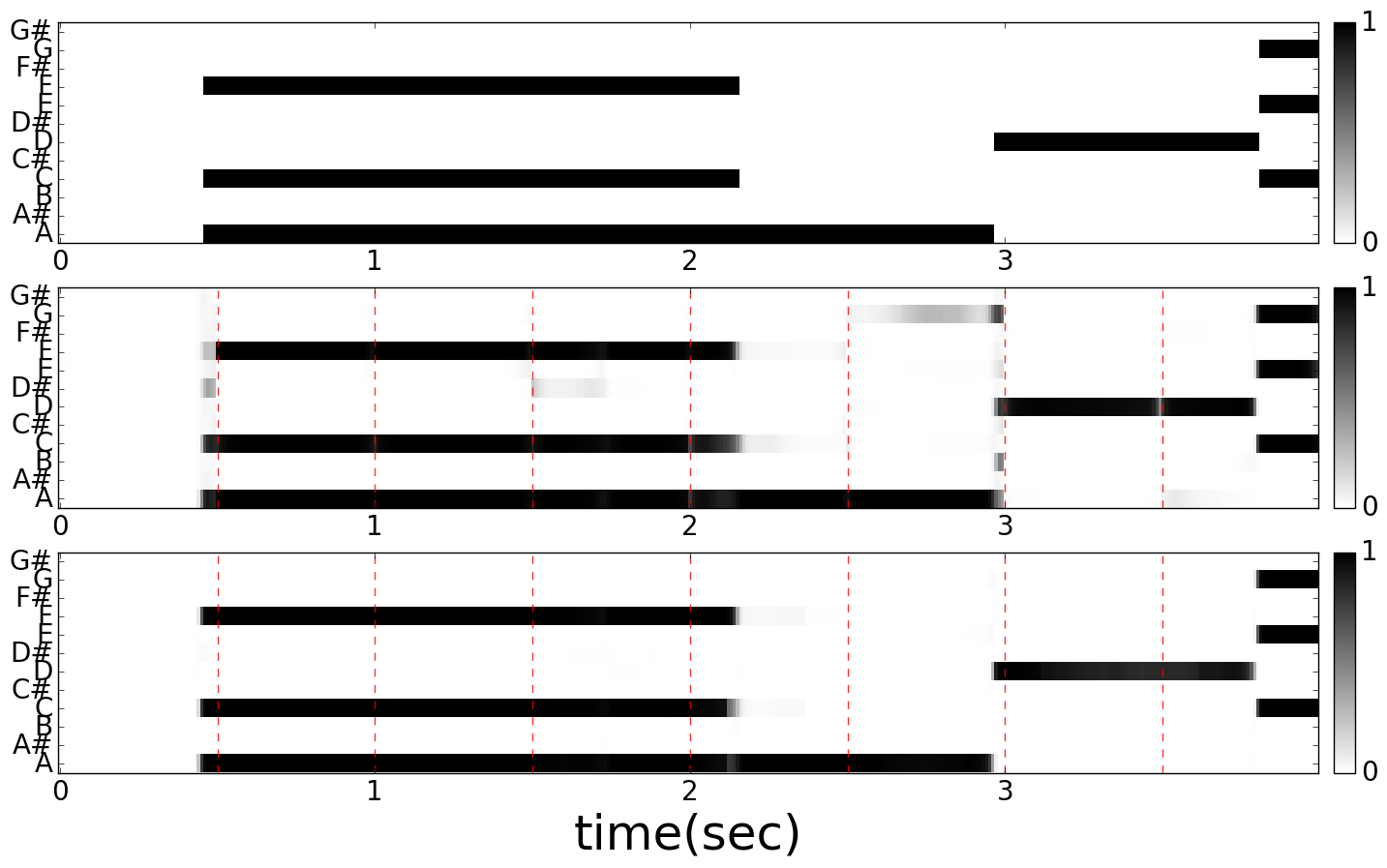}
\caption{Examples of bidirectional LSTM networks that predict 12 chroma: (a) ground truth, (b) without overlapping segmentation, (c) with overlapping segmentation. Dotted lines indicate the boundaries of segments}
\label{fig:overlap}
\end{figure}

To reduce the amount computation, our model works in sequence-to-sequence manner. In other words, the output of the network is a sequence with the same length of input segment.
Therefore, frames on the edges of a segment have only one-side context window.
We observed that contagious errors frequently occur on such frames as shown in Figure \ref{fig:overlap}b.
To tackle this problem we split the input sequence with 50\% overlapped segments and take only the middle part of output from each segment.
This procedure significantly increase transcription result as shown in Figure \ref{fig:overlap}c.

\subsubsection{Network Training}

\begin{figure}
\includegraphics[width=\columnwidth]{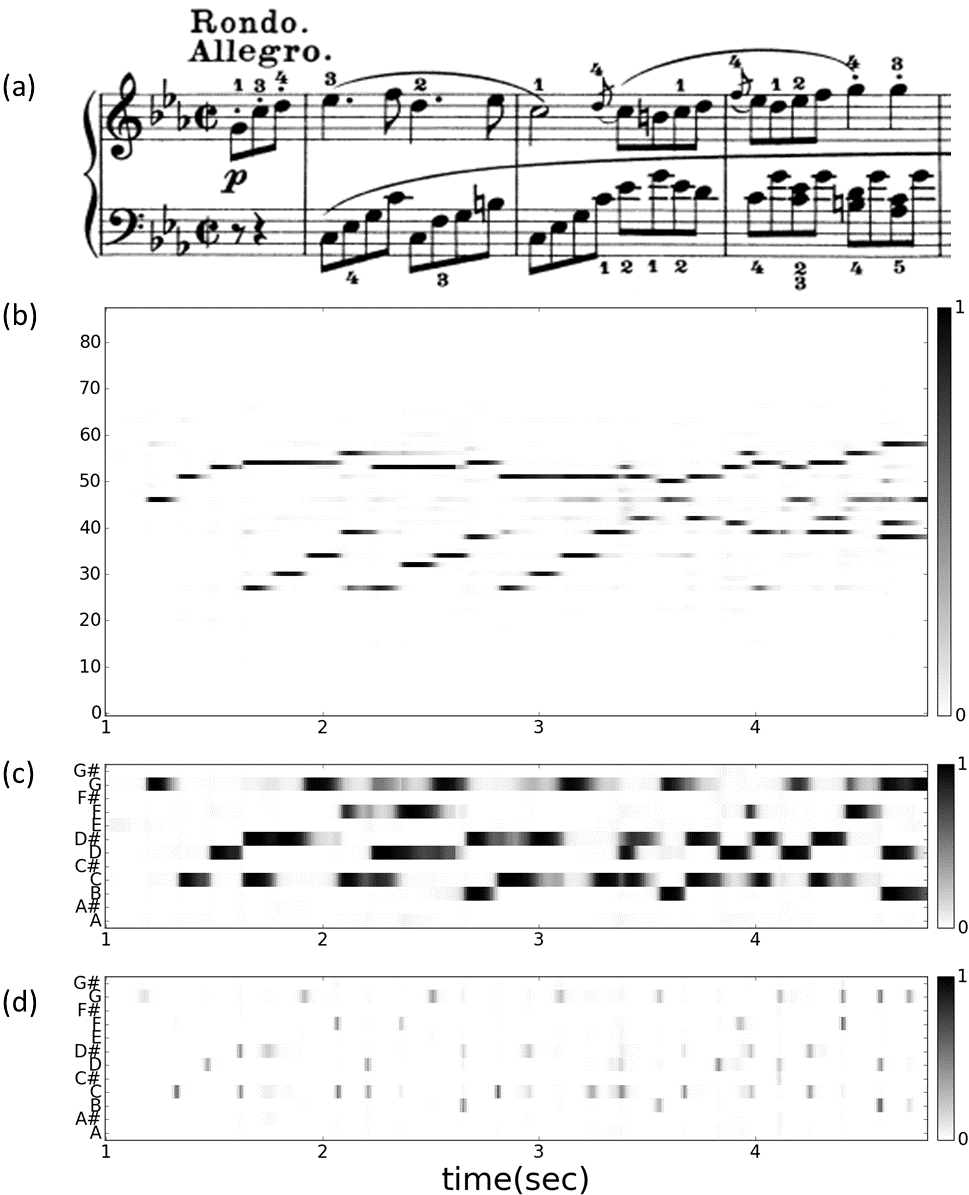}
\caption{(a) An excerpt of music score from Beethoven's 8th sonata. (b)-(d) the prediction outputs of the AMT systems: (b) 88 note (c) 12 chroma (d) 12 chroma onset.}
\label{fig:features}
\end{figure}

In order to train the networks, we used audio files and aligned MIDI files. 
The MIDI data was converted into a piano-roll representation with the same frame rate of the input filter-bank spectrogram (100 fps). For 88 notes and chroma labels, the elements of piano-roll representation were set to 1 between note onset and offset and otherwise to 0.
For chroma onset labels, only the elements that correspond to note onsets were set to 1.
The corresponding audio data was normalized with zero-mean and standard deviation of one over each filter in the training set.

We used dropout with a ratio of 0.5 and weight regularization with a value of $10^{-4}$ in each LSTM layer. This effectively improved the performance by generalization.
We used the network with stochastic gradient decent to minimize binary cross entropy loss function.
Learning rate was initially set as 0.1 and iteratively decreased by a factor of 3 when no improvement was observed for validation loss for 10 epochs (i.e. early stopping).
The training was stopped after six iterations.
Examples of the AMT outputs are presented in Figure \ref{fig:features}.
To verify the performance, frame-wise transcription performance for the 88-note AMT system was measured on the test sets. We used a fixed threshold of 0.5 to predict the note presence and measured the accuracy with F-score to make the results comparable to those in \cite{Kelz2016OnTranscription,Sigtia2016AnTranscription}. 
The resulting F-score was 0.7285 on average, which is better than the results of RNN with basic units \cite{Sigtia2016AnTranscription} and lower than those with fine-tuned frame-wise DNN and CNN \cite{Kelz2016OnTranscription}. 

\subsection{Alignment}
The AMT systems return two types of MIDI-level features. For chroma onset features, every onset was elongated for 10 frames (100 ms) with decaying weights of 1, $\sqrt{0.9}$, $\sqrt{0.8}$, ... , $\sqrt{0.1}$ as proposed in \cite{Ewert2009HighFeatures}.
The resulting features are combined by concatenation with either 88-note or 12-chroma AMT output features. The corresponding score MIDI was also converted into 88 note (or chroma) and the chroma onsets are elongated in the same manner before combined. 


We used euclidean distance to measure similarity between the two combined representations. We then applied the FastDTW algorithm \cite{Salvador2007FastDTWSpace} which is an approximate method to dynamic time warping (DTW).
FastDTW uses iterative multi-level approach with window constraints to reduce the complexity.
Because of the high frame rate of the features, it is necessary to employ low-cost algorithm.
While the original DTW algorithm has O($N^{2}$) time and space complexity, FastDTW operates in O($N$) complexity with almost the same accuracy.
M{\"u}ller et al. \cite{Muller2006AnSynchronization} also examined a similar multi-level DTW for the audio-to-score alignment task and reported similar results compared to the original DTW.
The radius parameter in the fastDTW algorithm, which defines the size of window to find an optimal path for each resolution refinement, was set to 10 in our experiment.

\section{Experiments}\label{sec:Experiments}

\subsection{Dataset}

We used the MAPS dataset \cite{Emiya2010MultipitchPrinciple}, specifically the `MUS' subset that contains large pieces of piano music, for training and evaluation.
Each piece consists of audio files and a ground-truth MIDI file. 
The audio files were rendered from the MIDI with nine settings of different pianos and recording conditions. This helped our model avoid overfitting to a specific piano tone.
The MIDI files served as the ground-truth annotation of the corresponding audio but some of them (ENSTDkCl and ENSTDkAm) are sometimes temporally inaccurate, which is more than 65 ms as described in \cite{Ewert2016PianoFramework}.

We conducted the experience with 4-fold cross validation with training and test splits from the configuration I\footnote{http://www.eecs.qmul.ac.uk/∼sss31/TASLP/info.html} in \cite{Sigtia2016AnTranscription}.
For each fold, 43 pieces were detached from the training set and used for validation.
As a result, each fold was composed of 173,  43  and 54 pieces for training, validation and test, respectively, as processed in \cite{Kelz2016OnTranscription}.


\subsection{Evaluation method}
In order to evaluate the audio-to-score alignment task, we need another MIDI representation (typically score MIDI) apart from the performance MIDI aligned with audio. We generated the separate MIDI by changing intervals between successive concurrent set of notes. Specifically, we multiplied a randomly selected value between 0.7 and 1.3 to modify the interval. This scheme of temporal distortion prevents the alignment path from being trivial and was also employed in previous work \cite{Ewert2009HighFeatures,muller2006efficient,joder2011conditional}.

After we obtained the alignment path through DTW, absolute temporal errors between estimated note onsets and ground truth were measured.
For each piece of music in the test set, mean value of the temporal errors and ratio of correctly aligned notes with varying thresholds were used to summarize the results. 

\subsection{Compared Algorithms}

To make a performance comparison, we reproduced two alignment algorithms proposed by Ewert et al. \cite{Ewert2009HighFeatures} and one by Carabias-Orti et al.\cite{Carabias-Orti2015AnWarping}. 
We performed the experiments with the same test set using the FastDTW algorithm but without any post-processing. 
Ewert's algorithms used a hand-crafted chromagram and onset features based on audio filter bank responses. 
Carabias-Orti's algorithm employed a non-negative matrix factorization to learn spectral basis of each note combination from spectrogram. 
The latter is designed only for audio-to-audio alignment while the former can be applied to both audio-to-audio and audio-to-MIDI alignment. Therefore, we made an audio version of the distorted MIDI using a high-quality sample-based piano synthesizer and employed it as an input. We tested Ewert's algorithms for both audio and MIDI cases.  
The temporal frame rate of features were adjusted to 100 fps for both algorithms. 

For the aligning task with Ewert's algorithms, we used the same FastDTW algorithm. 
But since the FastDTW algorithm cannot be directly applied to Carabias-Orti's algorithm due to its own distance calculation method, we applied a classic DTW algorithm, which employs an entire frame-wise distance matrix. 
Because of the limitation of memory, when reproducing Carabias-Orti's algorithm, we excluded 35 pieces that are longer than 400 seconds among the test sets.

Note that even though the dataset for evaluation is different, the results of two reproduced algorithms were similar to the results in their original works. The mean onset errors of Ewert's algorithm on piano music was 19 ms with 26 ms standard deviation \cite{Ewert2009HighFeatures}. The result introduced in the original Carabias-Orti's paper \cite{Carabias-Orti2015AnWarping} shows a quite large difference in terms of the mean of piecewise error, but we assumes that the difference is due to the change of the test set. The align rate of original result and our reproduced result were similar (50 ms: 74\% - 69\%, 100 ms: 90\% - 92\%, 200 ms: 95\% - 96\%). Hence, we assumed that our reproduction was reliable for the comparison.

\section{Results and Discussion}


\begin{table*}[tbp]
\def\arraystretch{1}
\centering
\resizebox{0.9\textwidth}{!}{%
\begin{tabular}{ll|c|c|c|c|c|c|c}
            						    &        & Mean   & Median & Std    & $\leq$ 10 ms & $\leq$	30 ms & $\leq$ 50 ms & $\leq$ 100 ms  \\ \hhline{=========}
\multicolumn{1}{l|}{\multirow{2}{*}{Proposed with onset}}    & chroma & 12.83  & 6.40   & 56.22  &  \textbf{92.01} & 97.44 & 98.31 & 98.98     \\ 
\multicolumn{1}{l|}{}    									& 88 note & \textbf{8.62}&\textbf{5.57} & 31.14  &91.60 & \textbf{98.00} & \textbf{98.97} & \textbf{99.61}  \\ \hline
\multicolumn{1}{l|}{\multirow{2}{*}{Proposed w/o onset}}     & chroma & 48.01  & 27.96  & 152.06 & 60.66 & 84.65 & 89.36 & 93.72        \\ 
\multicolumn{1}{l|}{} & 88 note    & 25.31  & 18.69  & 63.26  & 56.39 & 86.42 & 93.05 & 97.48\\ \hhline{==|=|=|=|=|=|=|=}
\multicolumn{2}{l|}{Ewert et. al. (audio-to-MIDI)}   		 & 16.44  & 13.64  & 32.52  & 71.78 & 91.38 & 95.50 & 98.03         \\ 
\multicolumn{2}{l|}{Ewert et. al. (audio-to-audio)}		 & 14.66  & 11.71  & \textbf{25.38}  & 71.53 & 92.43 & 96.91 & 99.13        \\
\multicolumn{2}{l|}{Carabias-Orti et. al.}        	     & 131.31 & 49.96  & 305.52 & 23.58 & 49.40 & 69.30 & 91.60         \\
\end{tabular}%
}
\caption{Results of the piecewise onset errors. Mean, median, and standard deviation of the errors are in millisecond. The right columns are the ratio of notes (\%) that are aligned within the onset error of 10 ms, 30 ms, 50 ms and 100 ms, respectively.}
\label{tab:result}
\end{table*}

\begin{figure} 
\includegraphics[width=\columnwidth]{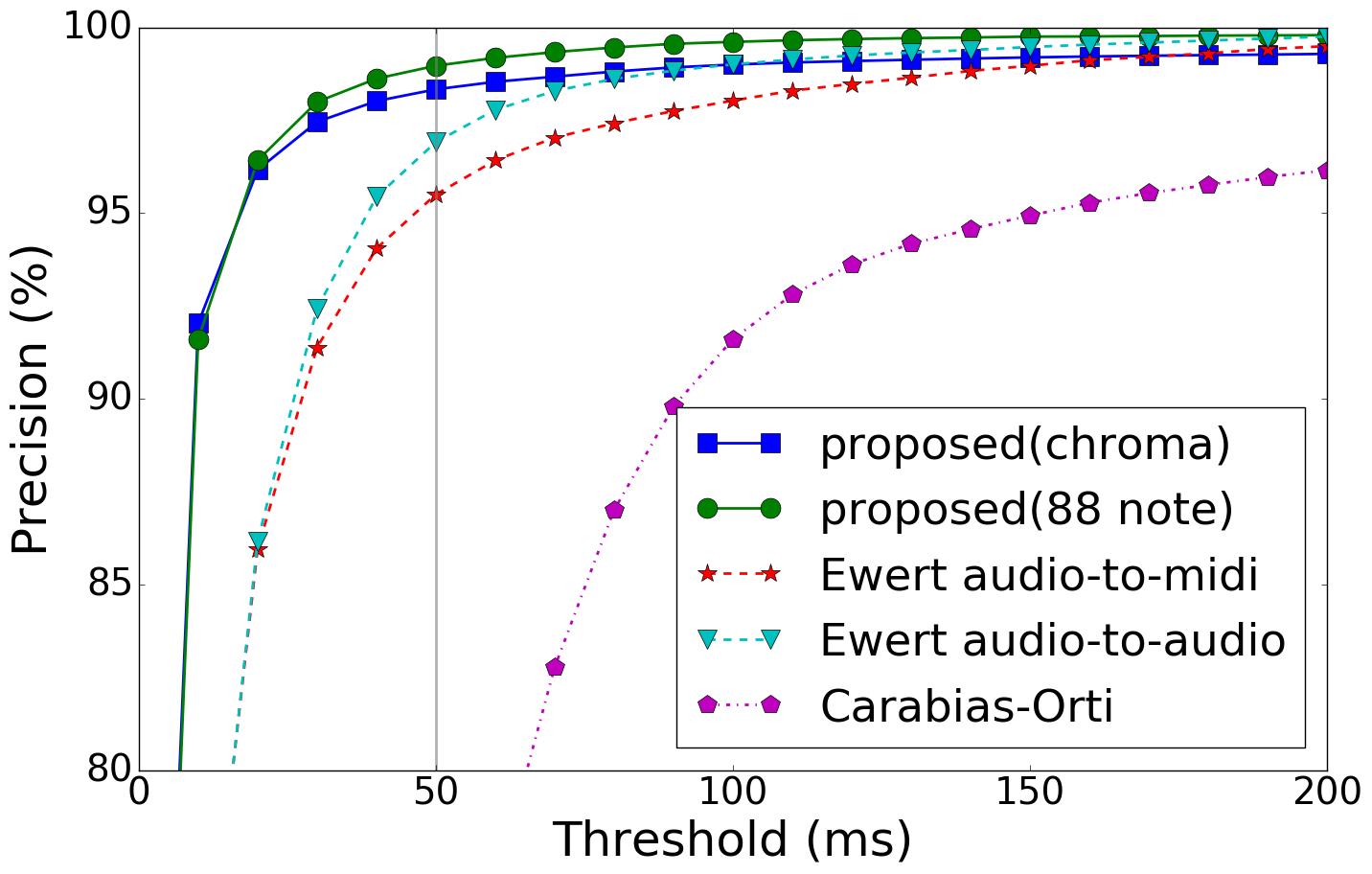}
\caption{Ratio of correctly aligned onsets as a function of threshold. Each point represent mean of piecewise precision. Some data points with lower than 80\% of precision are not shown in this figure.}
\label{fig:align_rate}
\end{figure}

\subsection{Comparison with Others}
Figure \ref{fig:align_rate} shows the results of the audio-to-score alignment from the compared algorithms. They represent the ratio of correctly aligned onsets in precision as a function of error threshold.
Typically, tolerance window with 50 ms is used for evaluation. 
However, because most of notes were aligned within 50 ms of temporal threshold, we varied the width tolerance window from 0 ms to 200 ms with 10 ms steps.

Overall, our proposed framework with 88 note combined with the chroma onsets achieved the best accuracy. Even with zero threshold, which means the best match with resolution of our system (10 ms), our proposed model with the 88-note output exactly aligned 52.55\% of notes. The ratio was increased to 91.60\% with 10 ms threshold.
The proposed framework using 12 chroma showed similar precision to the 88-note framework, but the accuracy was slightly lower.
Compared to Ewert's algorithms with hand-crafted features, our method shows significantly better performance especially in high resolution. 
Over 100 ms of threshold, our framework with chroma and Ewert's method shows similar precisions but the difference becomes significant with the intervals under 50 ms. 
Note that we penalized our framework compared to the audio-to-audio scenario of Ewert's algorithm because the audio-to-audio approach takes advantage from identical note velocities.
We suppose Ewert's algorithm performed better in the audio-to-audio scenario rather than the audio-to-MIDI for the same reason.
Carabias-Orti's algorithm shows lower precisions compared to others.
We assume that the difference mainly comes from the usage of onset features. 

For the fair comparison of the results, we should note that our framework is heavily dependent on the training set unlike the two other compared methods. 
 On the other hand, Carabias-Orti's  algorithm focused on dealing with various instruments and online alignment scenario, which were unable to be fully appreciated in our experiment.
\begin{figure} 
\includegraphics[width=\columnwidth]{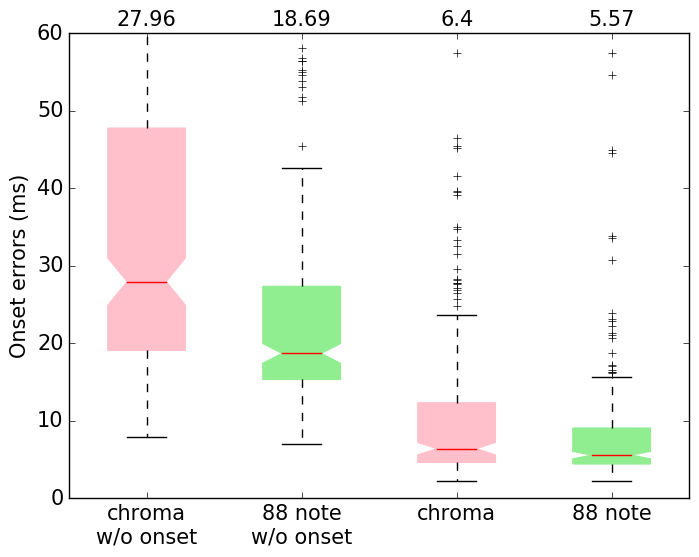}
\caption{Comparison of mean onset errors between models with/without chroma onset features. Each point corresponds to mean onset error of a piece. Outliers above 60 ms of errors are omitted in this figure. Each number on the top of the box indicates the median value in ms.}
\label{fig:mean_offset}
\end{figure}

\subsection{Effect of Chroma Onset Features}
On the second experiment, we further investigate the effect of chroma onset features.
We removed the onset features from each model and compared the mean onset errors and shows their distribution.
As can be seen in Figure \ref{fig:mean_offset}, the absence of onset features significantly decreases the performance. 
Thus we conclude that employing chroma onset features can compensate the limitation of normalized transcription features. As we stated in Section \ref{sec:introduction}, 
88 note representation shows much better results compared to those with chroma output features especially without onset.

Table \ref{tab:result} shows the statistics of piecewise onset errors. This shows that the use of chroma onset feature is crucial in our proposed method. The median of piecewise onset errors was decreased from 18.69 ms to 5.57 ms when applying chroma onset features to the 88-note system. The importance of the note onset feature for aligning piano music was also examined in \cite{Ewert2009HighFeatures}.


In addition to these experiments, we also aligned some real-world recordings through the trained system. Even though a quantitative evaluation has not been presented here, the sonification of the aligned MIDI files shows promising results. The  synchronized MIDI-audio examples are linked on our demo website\footnote{\url{http://mac.kaist.ac.kr/~ilcobo2/alignWithAMT}}.


\section{Conclusions}
In this paper, we proposed a framework for audio-to-score alignment of piano music using automatic music transcription. We built two AMT systems based on bidirectional LSTM that predict note existence and chroma onset. They provide MIDI-level features that can be compared with score MIDI to be used for the alignment algorithm. Our experiments with the MAPS dataset showed that the AMT-based features are effective in the alignment task and our proposed system outperforms compared approaches. The 88-note model with chroma onset worked best. We also showed that chroma onset features take a crucial role in improving the accuracy. In fact, the successful alignment performance might be possible because we used the same recording condition for both training and test sets. Considering this issue, we investigate the generalization capacity of our model by evaluating it on various datasets in the future. Also, we plan to improve the AMT system by using other types of deep neural networks.  


\begin{acknowledgments}
This work was supported by Korea Advanced Institute of
Science and Technology (project no. G04140049) and Korea Creative Content Agency (project no.  N04170044).
\end{acknowledgments} 

\bibliography{smc2017template}

\begin{thebibliography}{10}
\providecommand{\url}[1]{#1}
\csname url@samestyle\endcsname
\providecommand{\newblock}{\relax}
\providecommand{\bibinfo}[2]{#2}
\providecommand{\BIBentrySTDinterwordspacing}{\spaceskip=0pt\relax}
\providecommand{\BIBentryALTinterwordstretchfactor}{4}
\providecommand{\BIBentryALTinterwordspacing}{\spaceskip=\fontdimen2\font plus
\BIBentryALTinterwordstretchfactor\fontdimen3\font minus
  \fontdimen4\font\relax}
\providecommand{\BIBforeignlanguage}[2]{{%
\expandafter\ifx\csname l@#1\endcsname\relax
\typeout{** WARNING: IEEEtran.bst: No hyphenation pattern has been}%
\typeout{** loaded for the language `#1'. Using the pattern for}%
\typeout{** the default language instead.}%
\else
\language=\csname l@#1\endcsname
\fi
#2}}
\providecommand{\BIBdecl}{\relax}
\BIBdecl

\bibitem{Arzt2008AutomaticListening}
A.~Arzt, G.~Widmer, and S.~Dixon, ``{Automatic Page Turning for Musicians via
  Real-Time Machine Listening},'' in \emph{Proceedings of the conference on
  ECAI 2008: 18th European Conference on Artificial Intelligence}, vol.~1,
  no.~1, 2008, pp. 241--245.

\bibitem{Dannenberg2006MusicAccompaniment}
R.~B. Dannenberg and C.~Raphael, ``Music score alignment and computer
  accompaniment,'' \emph{Communications of the ACM}, vol.~49, no.~8, pp.
  38--43, 2006.

\bibitem{Widmer2003InFactor}
G.~Widmer, S.~Dixon, W.~Goebl, E.~Pampalk, and A.~Tobudic, ``{In Search of the
  Horowitz Factor},'' \emph{AI Magazine}, vol.~24, no.~3, pp. 111--130, 2003.

\bibitem{friberg1993perception}
A.~Friberg and J.~Sundberg, ``Perception of just-noticeable time displacement
  of a tone presented in a metrical sequence at different tempos,'' \emph{The
  Journal of The Acoustical Society of America}, vol.~94, no.~3, pp.
  1859--1859, 1993.

\bibitem{Dixon2005MATCH:Chest}
S.~Dixon and G.~Widmer, ``{MATCH: a music alignment tool chest},'' in
  \emph{Proceedings of the International Society for Music Information
  Retrieval Conference (ISMIR)}, 2005, pp. 492--497.

\bibitem{Ewert2009HighFeatures}
S.~Ewert, M.~M{\"{u}}ller, and P.~Grosche, ``{High resolution audio
  synchronization using chroma onset features},'' in \emph{Proc. IEEE
  International Conference on Acoustics, Speech and Signal Processing
  (ICASSP)}, 2009, pp. 1869--1872.

\bibitem{Arzt2014TheCompanion}
A.~Arzt, S.~B{\"{o}}ck, S.~Flossmann, H.~Frostel, M.~Gasser, C.~C.~S. Liem, and
  G.~Widmer, ``{The piano music companion},'' \emph{Frontiers in Artificial
  Intelligence and Applications}, vol. 263, no.~1, pp. 1221--1222, 2014.

\bibitem{Bock2012PolyphonicNetworks}
S.~B{\"o}ck and M.~Schedl, ``Polyphonic piano note transcription with recurrent
  neural networks,'' in \emph{Proc. IEEE International Conference on Acoustics,
  Speech and Signal Processing (ICASSP)}.\hskip 1em plus 0.5em minus
  0.4em\relax IEEE, 2012, pp. 121--124.

\bibitem{Williams1990AnTrajectories}
R.~J. Williams and J.~Peng, ``{An Efficient Gradient-Based Algorithm for
  On-Line Training of Recurrent Network Trajectories},'' \emph{Appears in
  Neural Computation}, no.~2, pp. 490--501, 1990.

\bibitem{Kelz2016OnTranscription}
R.~Kelz, M.~Dorfer, F.~Korzeniowski, S.~B{\"o}ck, A.~Arzt, and G.~Widmer, ``{On
  the Potential of Simple Framewise Approaches to Piano Transcription},'' in
  \emph{Proceedings of the International Conference on Music Information
  Retrieval (ISMIR)}, 2016, pp. 475--481.

\bibitem{Sigtia2016AnTranscription}
S.~Sigtia, E.~Benetos, and S.~Dixon, ``An end-to-end neural network for
  polyphonic piano music transcription,'' \emph{IEEE/ACM Transactions on Audio,
  Speech and Language Processing (TASLP)}, vol.~24, no.~5, pp. 927--939, 2016.

\bibitem{Salvador2007FastDTWSpace}
S.~Salvador and P.~Chan, ``{FastDTW : Toward Accurate Dynamic Time Warping in
  Linear Time and Space},'' \emph{Intelligent Data Analysis}, vol.~11, pp.
  561--580, 2007.

\bibitem{Muller2006AnSynchronization}
M.~M{\"{u}}ller, H.~Mattes, and F.~Kurth, ``{An efficient multiscale approach
  to audio synchronization},'' in \emph{Proc. International Conference on Music
  Informa- tion Retrieval (ISMIR)}, 2006, p. 192–197.

\bibitem{Emiya2010MultipitchPrinciple}
V.~Emiya, R.~Badeau, and B.~David, ``{Multipitch estimation of piano sounds
  using a new probabilistic spectral smoothness principle},'' \emph{IEEE
  Transactions on Audio, Speech and Language Processing}, vol.~18, no.~6, pp.
  1643--1654, 2010.

\bibitem{Ewert2016PianoFramework}
S.~Ewert and M.~Sandler, ``{Piano Transcription in the Studio Using an
  Extensible Alternating Directions Framework},'' vol.~24, no.~11, pp.
  1983--1997, 2016.

\bibitem{muller2006efficient}
M.~M{\"u}ller, H.~Mattes, and F.~Kurth, ``An efficient multiscale approach to
  audio synchronization.'' in \emph{Proc. International Conference on Music
  Informa- tion Retrieval (ISMIR)}.\hskip 1em plus 0.5em minus 0.4em\relax
  Citeseer, 2006, pp. 192--197.

\bibitem{joder2011conditional}
C.~Joder, S.~Essid, and G.~Richard, ``A conditional random field framework for
  robust and scalable audio-to-score matching,'' \emph{IEEE Transactions on
  Audio, Speech, and Language Processing}, vol.~19, no.~8, pp. 2385--2397,
  2011.

\bibitem{Carabias-Orti2015AnWarping}
J.~J. Carabias-Orti, F.~J. Rodr{\'\i}guez-Serrano, P.~Vera-Candeas,
  N.~Ruiz-Reyes, and F.~J. Ca{\~n}adas-Quesada, ``An audio to score alignment
  framework using spectral factorization and dynamic time warping.'' in
  \emph{Proc. International Conference on Music Informa- tion Retrieval
  (ISMIR)}, 2015, pp. 742--748.

\end{thebibliography}

\end{document}